\documentclass[journal,onecolumn]{IEEEtran}
\usepackage[dvips]{graphicx}
\usepackage[latin1]{inputenc}
\usepackage{amsmath}
\usepackage{times}
\usepackage{amsbsy,color}
\usepackage{latexsym}
\usepackage{amssymb}
\usepackage{graphicx}
\usepackage{subfigure}
\usepackage{amsfonts}
\usepackage{epsfig,subfigure,tikz}

\begin{document}

\title{Study of Distributed Spectrum Estimation Using Alternating Mixed Discrete-Continuous Adaptation}

\author{}
\author{ Rodrigo C. de Lamare\\

}

\maketitle

\begin{abstract}
This paper proposes a distributed alternating mixed
discrete-continuous (DAMDC) algorithm to approach the oracle
algorithm based on the diffusion strategy for parameter and spectrum
estimation over sensor networks. A least mean squares (LMS) type
algorithm that obtains the oracle matrix adaptively is developed and
compared with the existing sparsity-aware and conventional
algorithms. The proposed algorithm exhibits improved performance in
terms of mean square deviation and power spectrum estimation
accuracy. Numerical results show that the DAMDC algorithm achieves
excellent performance.
\end{abstract}

\begin{keywords}
Distributed processing, spectrum estimation, oracle Algorithm,
diffusion-LMS, sparsity-aware algorithms.
\end{keywords}

\section{Introduction}

\IEEEPARstart{D}{istributed} signal processing strategies are very
promising tools for solving parameter estimation problems in
wireless networks and applications such as sensor networks
\cite{lopes,cattivelli,mateos}. These techniques can exploit the
spatial diversity available in a network of sensors to obtain
increased estimation accuracy and robustness against sensor
failures.

Another set of tools for enhancing the performance of signal
processing algorithms is the exploitation of sparsity, work on which
initially dealt with centralized problems
\cite{candes,gu,delamarespl1,delamarespl07,yilun,jidf,fa10,eksioglu,angelosante,kalouptsidis,saalt,zhaocheng,zhaocheng2}
and, more recently, has examined distributed techniques
\cite{chouvardas,lorenzo1,lorenzo2,lorenzo3,arablouei,liu,dcg,dce,dta_ls,dcg_iet}
in several applications. A common strategy among the techniques
reported so far is the development of adaptive algorithms such as
the least mean squares (LMS)
\cite{yilun,saalt,lorenzo1,lorenzo2,lorenzo3,arablouei,dce,dta_ls}
and recursive least-squares (RLS)
\cite{eksioglu,angelosante,liu,dta_ls} using different penalty
functions. Such penalty functions perform a regularization that
attracts to zero the elements of the parameter vector with small
magnitudes. The most well-known and successful penalty functions are
the $l_{0}$-norm \cite{gu,eksioglu}, the $l_{1}$-norm \cite{yilun}
and the log-sum penalty \cite{candes,yilun}. The optimal algorithm
for processing sparse signals is known as the oracle algorithm
\cite{saalt}, which requires an exhaustive search for the location
of the non-zero coefficients followed by parameter estimation.

With the development and increasing deployment of mobile networks,
the frequency spectrum has become a resource that should be
exploited in a judicious way to avoid interference. By estimating
the power spectrum with spatially distributed sensors this resource
can be planned and properly exploited
\cite{lorenzo1,lorenzo2,lorenzo4}. Diffusion adaptation strategies
incorporating sparsity constraints have been used to solve
distributed spectrum estimation problems in \cite{lorenzo1} and
\cite{lorenzo2}. However, prior work on distributed techniques that
approach the oracle algorithm is rather limited, and adaptive
techniques that exploit potential sparsity of signals using discrete
and continuous variables have not been developed so far.

In this work, we propose a sparsity-aware distributed alternating
mixed discrete-continuous LMS (DAMDC-LMS) algorithm based on the
diffusion adapt-then-combine (ATC) protocol. We consider an
alternating optimization strategy with an LMS-type recursion along
with a mapping from continuous to discrete variables, which is used
to find the actual non-zero values, and another LMS-type recursion
that performs continuous adaptation. In particular, the proposed
DAMDC-LMS algorithm is incorporated into a distributed spectrum
estimation strategy. DAMDC-LMS is compared with prior art in a
distributed spectrum estimation application.

This paper is organized as follows. Section II describes the system
model and the problem statement. Section III presents the proposed
DAMDC-LMS algorithm. Section IV details the proposed algorithm for
an application to spectrum estimation. Section V presents and
discusses the simulation results. Finally, Section VI provides our
conclusions.

{\it Notation}: In this paper, matrices and vectors are designated
by boldface upper case letters and boldface lower case letters,
respectively. The superscript $(.)^{H}$ denotes the Hermitian
operator, $\|.\|^{1}$ refers to the $l_{1}$-norm and $E[\cdot]$
denotes expected value.

\section{System Model and Problem Statement}

\begin{figure}[hbt]
\centering
\includegraphics[scale=0.35]{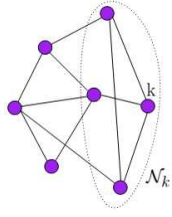}\vspace{-0.75em}
\caption{Network topology with $N$ nodes.} \label{1}
\end{figure}

We consider a network that is partially connected and consists of
$N$ nodes that exchange information among themselves. Each node $k$
employs a parameter estimator and has its neighborhood described by
the set ${\mathcal N}_{k}$, as shown in Fig. \ref{1}. The task of
parameter estimation is to adjust an \textit{M} $\times 1$ weight
vector $\boldsymbol\omega_{k,i}$ at each node $k$ and time $i$ based
on an \textit{M}$\times 1$ input signal vector $\boldsymbol x_{k,i}$
and ultimately estimate an unknown \textit{M} $\times 1$ system
parameter vector $\boldsymbol\omega_{0}$ \cite{lopes}. The desired
signal $d_{k,i}$ at each time $i$ and node $k$ is drawn from a
random process and given by
\begin{equation}
\ d_{k,i}=\boldsymbol\omega_{0}^{H}\boldsymbol x_{k,i}+n_{k,i},
\end{equation}
where $n_{k,i}$ is measurement noise.

We consider a distributed estimation problem for a network in which
each agent $k$ has access at each time instant to a realization of
zero-mean spatial data $\{d_{k,i} , \boldsymbol x_{k,i}\}$
\cite{lopes,mateos}. The goal of the network is to minimize the
following cost function:
\begin{equation}
\label{Eqn4:cost_function}
\begin{split}
C(\boldsymbol\omega_{k,i}) &
=\sum_{k=1}^NE[|d_{k,i}-\hat{d}_{k,i}|^{2}] \\ & =\sum_{k=1}^N
E[|d_{k,i} - \boldsymbol\omega_{k,i}^{H}\boldsymbol x_{k,i}|^{2}],~
{\rm for}~ k = 1, 2, \ldots, N,
\end{split}
\end{equation}
By solving this minimization problem one can obtain the optimum
solution for the weight vector at each node. For a network with
possibly sparse parameter vectors, the cost function might also
involve a penalty function that exploits sparsity. In what follows,
we present a novel distributed diffusion technique to approach the
oracle algorithm and efficiently solve (\ref{Eqn4:cost_function})
under sparseness conditions.

\section{Proposed DAMDC-LMS Algorithm}

\begin{figure}[!htb]
\begin{center}
\def\epsfsize#1#2{0.85\columnwidth}
\epsfbox{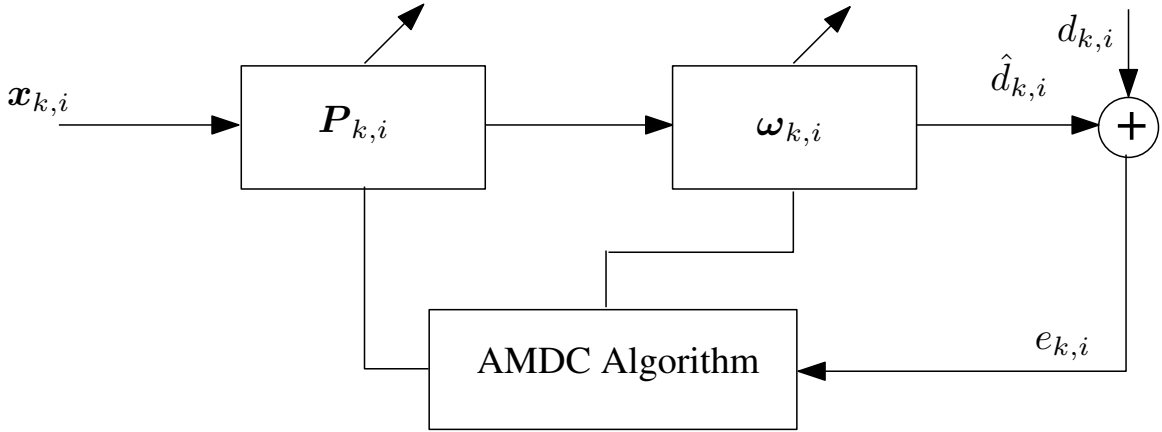} \vspace{-0.75em} \caption{ Proposed adaptive
scheme at node $k$.} \label{2}
\end{center}
\end{figure}

In this section, we detail the proposed distributed scheme and
DAMDC-LMS algorithm using the diffusion ATC strategy. The proposed
scheme for each agent $k$ of the network is shown in Fig. \ref{2}.
The output estimate of the proposed scheme is given by
\begin{equation}
\begin{split}
\hat{d}_{k,i} & = \boldsymbol\omega_{k,i}^{H}\boldsymbol
P_{k,i}\boldsymbol x_{k,i}  = \boldsymbol p^{T}_{k,i}\boldsymbol
W^{*}_{k,i}\boldsymbol x_{k,i}\\ & =\boldsymbol
x^{T}_{k,i}\boldsymbol W^{*}_{k,i}\boldsymbol p_{k,i}=\boldsymbol
x^{T}_{k,i}\boldsymbol P_{k,i}\boldsymbol \omega^{*}_{k,i},
\label{relation}
\end{split}
\end{equation}
where the parameter vector ${\boldsymbol \omega}_{k,i}$ is a column
vector of $M$ coefficients related to the diagonal matrix $
\boldsymbol W_{k,i} = {\rm diag}(\boldsymbol\omega_{k,i})$. The
matrix $\boldsymbol P_{k,i}$ is a square diagonal matrix with $M$
elements that is applied to the input vector $\boldsymbol x_{k,i}$
and aims to simulate the oracle algorithm by identifying the null
positions of $\boldsymbol\omega_{0}$.

In order to obtain recursions for ${\boldsymbol P}_{k,i}$ and
${\boldsymbol \omega}_{k,i}$ we compute the stochastic gradient of
the cost function in (\ref{Eqn4:cost_function}) with respect to both
parameters, where the optimization of ${\boldsymbol P}_{k,i}$
involves discrete variables and ${\boldsymbol \omega}_{k,i}$ deals
with continuous variables. In particular, we develop an alternating
optimization approach using an LMS type algorithm that consists of a
recursion for ${\boldsymbol P}_{k,i}$ and another recursion for
${\boldsymbol \omega}_{k,i}$ that are employed in an alternating
fashion.



 {In order to compute ${\boldsymbol P}_{k,i}$ and
${\boldsymbol \omega}_{k,i}$ we must solve the mixed
discrete-continuous non-convex optimization problem:
\begin{equation}
\begin{split}
{\boldsymbol p}_{k,i}^*, {\boldsymbol \omega}_{k,i}^* & =
\min_{{\boldsymbol p}_{k,i} \in {\mathcal I}^{M \times 1},~~
{\boldsymbol \omega}_{k,i}\in {\mathcal C}^{M \times 1}}
C({\boldsymbol p}_{k,i}, {\boldsymbol \omega}_{k,i}),\\ & ~{\rm
for}~ k = 1, 2, \ldots, N, \label{mdc_problem}
\end{split}
\end{equation}
where
\begin{equation}
C({\boldsymbol p}_{k,i}, {\boldsymbol \omega}_{k,i})
 =\sum_{k=1}^N E[|d_{k,i}-\boldsymbol p^{T}_{k,i}{\boldsymbol
W}^{H}_{k,i}{\boldsymbol x}_{k,i}|^{2}], \label{Eqn5:MSE_1}
\end{equation}
${\boldsymbol p}_{k,i}$ contains the elements of the main diagonal
of ${\bf P}_{k,i}$, and ${\mathcal I}^{M \times 1}$ denotes the set
of $M$-dimensional
binary vectors with values $0$ and $1$. 
Since the problem in (\ref{mdc_problem}) is NP-hard, we
resort to an approach that assumes ${\boldsymbol p}_{k,i}$ is a
real-valued continuous parameter vector for its computation and then
map ${\boldsymbol p}_{k,i}$ to discrete values.} The relations in
(\ref{relation}) allow us to compute the gradient of the cost
function with respect to ${\boldsymbol p}_{k,i}$ and ${\boldsymbol
\omega}_{k,i}$ and their diagonal versions ${\boldsymbol P}_{k,i}$
and ${\boldsymbol W}^{H}_{k,i}$, respectively. The gradient of the
cost function with respect to ${\boldsymbol p}_{k,i}$ is given by
\begin{equation}
\label{Eqn7:MSE_derivation}
\begin{split}
 \nabla_{\boldsymbol p_{k,i}}C(\boldsymbol
p_{k,i},\boldsymbol\omega_{k,i})&
=\frac{\partial}{\partial\boldsymbol
p_{k,i}}\Big(E|d_{k,i}|^{2}-(\boldsymbol p^{T}_{k,i}\boldsymbol
W^{*}_{k,i}E[d^{*}_{k,i}\boldsymbol x_{k,i}])\\
& \quad +\boldsymbol p^{T}_{k,i}\boldsymbol W^{*}_{k,i}E[\boldsymbol
x_{k,i}\boldsymbol x^{H}_{k,i}\boldsymbol W_{k,i}\boldsymbol
p_{k,i}] \\ & \quad -E[d_{k,i}\boldsymbol x^{H}_{k,i}]\boldsymbol
W_{k,i}\boldsymbol p_{k,i} \Big).
\end{split}
\end{equation}
Replacing the expected values with instantaneous values, we obtain
\begin{equation}\label{Eqn8:MSE_derivation2}
\begin{split}
\hat{\nabla}_{\boldsymbol p_{k,i}}C(\boldsymbol
p_{k,i},\boldsymbol\omega_{k,i}) &
=\frac{\partial}{\partial\boldsymbol
p_{k,i}}\Big(|d_{k,i}|^{2}-\boldsymbol p^{T}_{k,i}\boldsymbol
W^{*}_{k,i}d^{*}_{k,i}\boldsymbol x_{k,i}\\ & \quad +\boldsymbol
p^{T}_{k,i}\boldsymbol W^{*}_{k,i}\boldsymbol x_{k,i}\boldsymbol
x^{H}_{k,i}\boldsymbol W_{k,i}\boldsymbol p_{k,i} \\ & \quad
-d_{k,i}\boldsymbol x^{H}_{k,i}\boldsymbol W_{k,i}\boldsymbol
p_{k,i}\Big).
\end{split}
\end{equation}
Computing the gradient of the cost function with respect to
${\boldsymbol p}_{k,i}$, we obtain
\begin{equation}\label{Eqn9:MSE_derivation3}
\begin{split}
\hat{\nabla}_{\boldsymbol p_{k,i}}C(\boldsymbol
p_{k,i},\boldsymbol\omega_{k,i}) & = d^{*}_{k,i}\boldsymbol
W^{*}_{k,i}{\boldsymbol x}_{k,i}-d_{k,i}\boldsymbol
W^{T}_{k,i}\boldsymbol x^{*}_{k,i}\\
 & \quad + \boldsymbol W^{*}_{k,i}\boldsymbol x_{k,i}\boldsymbol x^{H}_{k,i}\boldsymbol W_{k,i} \boldsymbol
 p_{k,i} \\
 & \quad  + \boldsymbol W^{T}_{k,i}\boldsymbol x^{*}_{k,i}\boldsymbol x^{T}_{k,i}\boldsymbol W^{H}_{k,i}\boldsymbol p_{k,i}.
\end{split}
\end{equation}
Grouping common terms, we arrive at
\begin{equation}\label{Eqn10:Grouping}
\begin{split}
\hat{\nabla}_{\boldsymbol p_{k,i}}C(\boldsymbol
p_{k,i},\boldsymbol\omega_{k,i}) & = -\Big(d_{k,i}- \boldsymbol
x^{H}_{k,i} \boldsymbol W_{k,i}\boldsymbol p_{k,i}\boldsymbol
W_{k,i}\boldsymbol x_{k,i}\\
& \quad  +d_{k,i}- \boldsymbol x^{T}_{k,i}\boldsymbol
W^{H}_{k,i}\boldsymbol p_{k,i}\boldsymbol W^{T}_{k,i}\boldsymbol
 x^{*}_{k,i}\Big),
\end{split}
\end{equation}
where $\boldsymbol p_{k,i}$ is a real parameter vector, $\boldsymbol
p_{k,i}=\boldsymbol p^{*}_{k,i}$, $ \boldsymbol
p^{H}_{k,i}=[\boldsymbol p^{*}_{k,i}]^{T}=\boldsymbol p^{T}_{k,i}$.
Since $\boldsymbol W_{k,i}$ is symmetric, i.e., $ \boldsymbol
W^{T}_{k,i}=\boldsymbol W_{k,i}$, we have $ \boldsymbol
W^{H}_{k,i}=[\boldsymbol W^{*}_{k,i}]^{T}=[\boldsymbol
W^{T}_{k,i}]^{*}=\boldsymbol W^{*}_{k,i}$. The terms in
(\ref{Eqn10:Grouping}) represent the sum of a vector and its
conjugate:
\begin{equation}
\label{Eqn11:Sum}
\begin{split}
\hat{\nabla}_{\boldsymbol p_{k,i}} C(\boldsymbol
p_{k,i},\boldsymbol\omega_{k,i}) & = -\underbrace{\Big(\big(d_{k,i}-
\boldsymbol x^{H}_{k,i}\boldsymbol W_{k,i}\boldsymbol
p_{k,i}\big)\boldsymbol W^{*}_{k,i}\boldsymbol x_{k,i}}_{A}\\
& \quad +\underbrace{\big(d_{k,i}- \boldsymbol
x^{T}_{k,i}\boldsymbol W^{H}_{k,i}\boldsymbol
p_{k,i}\big)\boldsymbol W_{k,i}\boldsymbol
x^{*}_{k,i}\Big)}_{A^{*}}.
\end{split}
\end{equation}
Applying the property $A + A^{*}=2\Re(A)$, we have
\begin{equation}\label{Eqn12:Sum_Conj}
\hat{\nabla}_{\boldsymbol p_{k,i}}\mbox{MSE}(\boldsymbol
p_{k,i},\boldsymbol\omega_{k,i})=2\Re(A).
\end{equation}
The recursion to update the parameter vector ${\boldsymbol p}_{k,i}$
is given by  {
\begin{equation}
\begin{split}
\label{Eqn13:P_recursion} \boldsymbol p_{k,i+1} & = \boldsymbol
p_{k,i} -\eta\hat{\nabla}_{\boldsymbol
p_{k,i}}\mbox{MSE}(\boldsymbol p_{k,i},\boldsymbol W_{k,i})\\
& =\boldsymbol p_{k,i} + 2\eta\Re(e_{p_{k,i}}^{*}\boldsymbol
x^{H}_{k,i}\boldsymbol W_{k,i}),
\end{split}
\end{equation}}
where the error signal is given by
\begin{equation}
\label{Eqn14:error}
\ e_{p_{k,i}}=d_{k,i}- \boldsymbol p^{T}_{k,i}\boldsymbol
W_{i-1}\boldsymbol x_{k,i}.
\end{equation}
For the update of the parameter vector ${\boldsymbol w}_{k,i}$, we
can apply well-known adaptive algorithms. By computing the gradient
of the cost function with respect to ${\boldsymbol w}_{k,i}^*$, we
have
\begin{equation} \label{eqn16new}
\nabla C_{{\boldsymbol w}^*_{k,i}}({\boldsymbol p}_{k,i},
{\boldsymbol w}_{k,i}) = (d_{k,i} - \boldsymbol
x^{T}_{k,i}{\boldsymbol P}_{k,i}
\boldsymbol\omega^{*}_{i-1})^{*}{\boldsymbol P}_{k,i}{\boldsymbol
x}_{k,i}
\end{equation}
The following LMS type recursion updates the parameter vector
${\boldsymbol \omega}_{k,i}$:
\begin{equation}
\label{Eqn17:W_Recursion2}
\boldsymbol\omega_{k,i+1}=\boldsymbol\omega_{k,i}+\mu
e^{*}_{k,i}\boldsymbol P_{k,i}\boldsymbol x_{k,i},
\end{equation}
where the error signal is given by $e_{k,i} = d_{k,i}-\boldsymbol
x^{T}_{k,i}\boldsymbol P_{k,i}\boldsymbol\omega^{*}_{i-1}$.

The recursions for ${\boldsymbol p}_{k,i}$ and ${\boldsymbol
\omega}_{k,i}$ using the ATC protocol \cite{lopes,cattivelli}
 {for $k=1, 2, \ldots, N$ are then given by}
\begin{equation}
\label{Eqn18:Distributed_RecursionP} \boldsymbol
p_{k,i+1}=\boldsymbol p_{k,i} + 2\eta \Re(e_{p_{k,i}}^{*}\boldsymbol
x_{k,i}^{H}\boldsymbol W_{k,i}),
\end{equation}
\begin{equation}
\label{Eqn19:Distributed_Recursion}
\boldsymbol\varphi_{k,i+1}=\boldsymbol\omega_{k,i-1}+\mu e^{*}_{k,i}\boldsymbol P_{k,i}\boldsymbol x_{k,i},\\
\end{equation}
\begin{equation}
\label{Eqn20:Distributed_Combination} \boldsymbol
\omega_{k,i}=\sum_{l\in N_{k}}a_{lk}\boldsymbol\varphi_{l,i},
\end{equation}
where (\ref{Eqn18:Distributed_RecursionP}) and
(\ref{Eqn19:Distributed_Recursion}) are the adaptation step, and
(\ref{Eqn20:Distributed_Combination}) is the combination step of the
ATC protocol. The combining coefficients of the latter are
represented by $a_{lk}$ and should comply with
\begin{equation}
\label{Eqn21:combinig coefficients} \sum_{l \in N_{k}} a_{lk}=1,\ l
\in N_{k,i}, \forall k.
\end{equation}
The strategy adopted in this work for the $a_{lk}$ combiner is the
Metropolis rule \cite{lopes} given by
\begin{equation}\label{Eqn22:Metropolis rule}
a_{kl}=\left\{\begin{array}{ll} \frac{1}
{max\{|\mathcal{N}_k|,|\mathcal{N}_l|\}}\ \ $if\  $k\neq l$\ are linked$,\\
1 - \sum\limits_{l\in \mathcal{N}_k / k} a_{kl}, \ \ $for\  $k$\ =\
$l$$.
\end{array}
\right.
\end{equation}
In order to compute the discrete vector $\boldsymbol p_{k,i}$, we
rely on a simple approach that maps the continuous variables into
discrete variables, which is inspired by the likelihood ascent
approach adopted for detection problems in wireless communications
\cite{vardhan,las_li}. The initial value at each node is an all-one
vector ($\boldsymbol p_{k,0}= \boldsymbol 1$ or $\boldsymbol
P_{k,0}=\boldsymbol I$). The $\boldsymbol \omega_{k,i}$ vector is
initialized as an all-zero vector ($\boldsymbol\omega_{k,0}=
\boldsymbol 0$ or $\boldsymbol W_{k,0}= \boldsymbol 0$).
 {After each iteration of
(\ref{Eqn18:Distributed_RecursionP})}, we obtain discrete values
from $\boldsymbol p_{k,i}$ using the following rule for $m = 1,
\ldots, M$:
\begin{equation}\label{Eqn23:P rule}
 p_{k,i+1}^{m}=\left\{\begin{array}{ll}
1,  \ $if\  $ p_{k,i}^{m}> \tau, \\
0, \  \mbox{otherwise},
\end{array}
\right.
\end{equation}
 {where $\tau$ is a threshold used to determine the
positions of the non-zero values of the parameter vector
$\boldsymbol p_{k,i}$. The goal is to approach the results of the
oracle algorithm and an appropriate value for $\tau$ can be obtained
experimentally.}

\section{Distributed Spectrum Estimation using the DAMDC-LMS Algorithm}

We now illustrate the use of DAMDC-LMS in distributed spectrum
estimation, which aims to estimate the spectrum of a transmitted
signal $s$ with $N$ nodes using a wireless sensor network
\cite{mateos,lorenzo1,lorenzo2}. The power spectral density (PSD) of
the signal $s$ at each frequency denoted by $\Phi_{s}(f)$ is given
by
\begin{equation}\label{Eqn24:PSD}
\ \Phi_{s}(f)=\sum_{m=1}^{M}b_{m}(f)\omega_{0m}=\boldsymbol
b_{0}^{T}(f)\boldsymbol\omega_{0},
\end{equation}
where $\boldsymbol b_{0}(f)=[ b_{1}(f),...,b_{M}(f)]^{T}$ is the
vector of basis functions evaluated at frequency $f$,
$\boldsymbol\omega_{0}=[\omega_{01},...,\omega_{0M}]$ is a vector of
weighting coefficients representing the transmit power of the signal
$s$ over each basis, and $M$ is the number of basis functions. For
$M$ sufficiently large, the basis expansion in (\ref{Eqn24:PSD}) can
approximate well the spectrum. Possible choices for the set of basis
functions $\{{b_{m}(f)\}}_{m=1}^M$ include rectangular functions,
raised cosines, Gaussian bells and splines \cite{dcg_iet}.

We denote the channel transfer function between a transmit node
conveying the signal $s$ and receive node $k$ at time instant $i$ by
$H_{k}(f,i)$, and thus the PSD of the received signal observed by
node k can be expressed as
\begin{equation}\label{Eqn25:PSD2}
\begin{split}
\Phi_{k}(f,i) & =|H_{k}(f,i)|^{2} \Phi_{s}(f)+\upsilon^{2}_{n,k},\\
 & =\sum_{m=1}^{M}|H_{k}(f,i)|^{2}b_{m}(f)\omega_{0m}+
\upsilon^{2}_{n,k}, \\ & = \boldsymbol
b_{k,i}^{T}(f)\boldsymbol\omega_{0m}+ \upsilon^{2}_{n,k}.
\end{split}
\end{equation}
where $\boldsymbol
b_{k,i}^{T}(f)=[|H_{k}(f,i)|^{2}b_{m}(f)]^{M}_{m=1}$ and
$\upsilon_{n,k}^{2}$ is the noise power of the receiver at node $k$.

Following the distributed model, at every iteration $i$ every node
$k$ measures the PSD $\Phi_{k}(f,i)$ presented in (\ref{Eqn25:PSD2})
over $N_{c}$ frequency samples
$f_{j}=f_{min}:(f_{max}-f_{min})/N_{c}:f_{max}$, for $j = 1,...,
N_{c}$, the desired signal is given by
\begin{equation}\label{Eqn26:Desired_PSD}
\ d_{k,i}(j)=\boldsymbol
b_{k,i}^T(f_{j})\boldsymbol\omega_{0}+\upsilon_{n,k}^2 + n_{k,i}(j),
\end{equation}
where the last term denotes the observation noise with zero mean and
variance $\sigma_{n,j}^{2}$. The noise power $\upsilon_{n,k}^{2}$ at
the receiver of node $k$ can be estimated with high accuracy in a
preliminary step using, e.g., an energy estimator over an idle band,
and then subtracted from (\ref{Eqn26:Desired_PSD}). A linear model
is obtained from the measurements over $N_{c}$ contiguous channels
\begin{equation}\label{Eqn27:Linear_PSD}
\ \boldsymbol d_{k,i}=\boldsymbol B_{k,i}\boldsymbol\omega_{0}+
\boldsymbol n_{k,i},
\end{equation}
where $\boldsymbol B_{k,i}=[\boldsymbol
b_{k,i}^T(f_{j})]_{j=1}^{N_{c}}\in {\mathcal R}^{N_{c}\times M}$,
and $\boldsymbol n_{k,i}$ is a zero mean random vector. Then we can
introduce the cost function for each agent $k$ described by
 {
\begin{equation}\label{Eqn28:Distributed_CostFunction}
\ C(\boldsymbol\omega_{k,i})=E[|\boldsymbol d_{k,i}-\boldsymbol
B_{k,i}\boldsymbol\omega_{k,i}|^{2}],~ {\rm for}~ k = 1, \ldots, N.
\end{equation}}
Once we have the cost function, the DAMDC-LMS algorithm can be
applied by introducing the discrete parameter vector $\boldsymbol
p_{k,i}$ in (\ref{Eqn28:Distributed_CostFunction}), which results in
 {
\begin{equation}\label{Eqn29:Distributed_CostFunction2}
\ C(\boldsymbol\omega_{k,i},\boldsymbol p_{k,i})=E[|\boldsymbol
d_{k,i}-\boldsymbol B_{k,i}\boldsymbol
P_{k,i}\boldsymbol\omega_{k,i}|^{2}],~ {\rm for}~ k = 1, \ldots, N,
\end{equation}
where} $\boldsymbol P_{k,i}$ is the $B\times B$ diagonal matrix  to
exploit the sparsity for a more accurate spectrum estimation.
Introducing the matrix $\boldsymbol P_{k,i}$ for exploiting sparsity
in the recursions (\ref{Eqn18:Distributed_RecursionP}),
(\ref{Eqn19:Distributed_Recursion}) and
(\ref{Eqn20:Distributed_Combination}), we obtain for $k = 1, 2,
\ldots, N$:
\begin{equation}\label{Eqn30:P_PSD}
{\rm Adaptation} \left\{\begin{array}{ll}
\boldsymbol p_{k,i+1}=\boldsymbol p_{k,i} + 2\eta \Re(e_{p_{k,i}}^{*}\boldsymbol B_{k,i}^{H}\boldsymbol W_{k,i-1})\\
p_{k,i+1}^{m}=\left\{\begin{array}{ll}
1,  \ $if\  $ p_{k,i}^{m}> \tau, ~{\rm for}~m = 1, \ldots, M,\\
0, \  \mbox{otherwise},
\end{array}
\right.\\
\boldsymbol\varphi_{k,i+1}=\boldsymbol\omega_{k,i-1}+\mu e^{*}_{k,i}\boldsymbol P_{k,i}\boldsymbol B_{k,i}, \\
\end{array}
\right.
\end{equation}
\begin{equation}\label{Eqn31:ATC2_PSD}
{\rm Combination} \left\{\begin{array}{ll} \boldsymbol
\omega_{k,i}=\sum_{l\in N_{k}}a_{lk}\boldsymbol
P_{k,i}\boldsymbol\varphi_{l,i}.
\end{array}
\right.
\end{equation}
The positions in $\boldsymbol p _{k,i}$ with ones indicate the
information content at each node and sample of the signal. With this
approach, we can identify the positions of the non-zero coefficients
of the frequency spectrum and achieve performance similar to that of
the oracle algorithm as seen in the following section.

\section{Simulation Results}

In this section, we evaluate the performance of the DAMDC-LMS
algorithm for distributed spectrum estimation using sensor networks,
where DAMDC-LMS is compared with existing algorithms. The results
are shown in terms of the mean square deviation (MSD), power and PSD
estimation.

\begin{figure}[!htb]
\begin{center}
\def\epsfsize#1#2{0.85\columnwidth}
\epsfbox{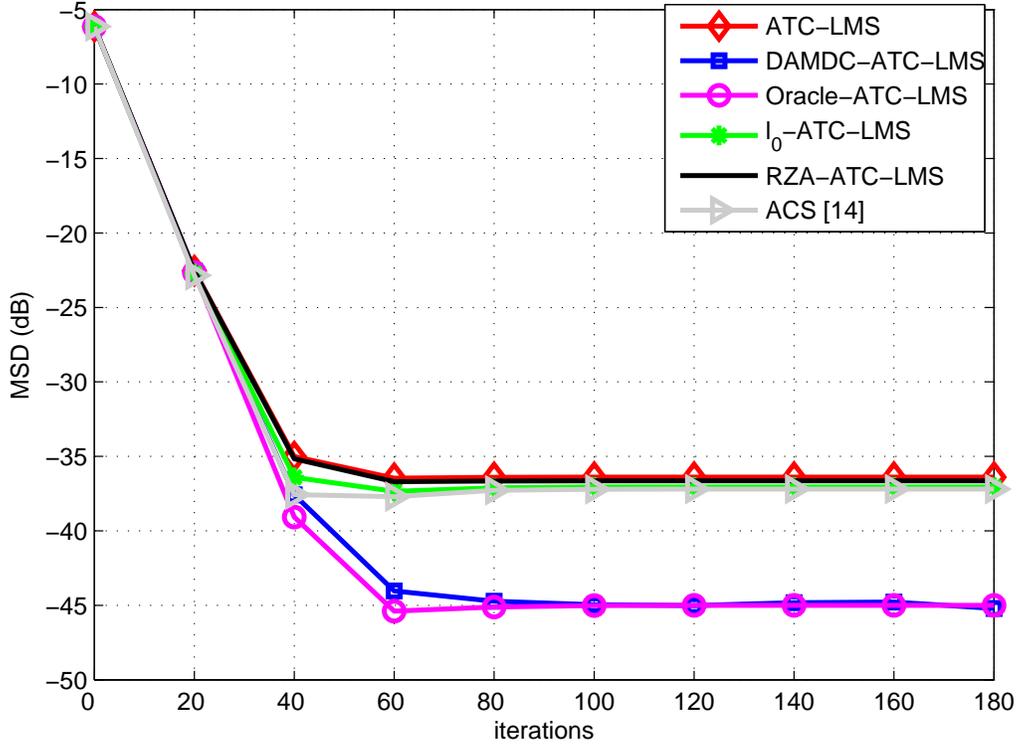} \vspace{-1.3em} \caption{ MSD for distributed
spectrum estimation. Parameters: $\mu = 0.05$, $\eta = 5 × 10^{-2}$
, $\rho_{\rm RZA} = 3.5 × 10^{-5}$, $\rho_{l_0} = 3.5 × 10^{-5}$,
$\rho_{ACS}= 10^{-3}$, $\beta = 5$, $\beta_ACS= 50$, $\epsilon =
0.1$, ${\rm SNR} = 30dB$, $\tau=1$ and $S = 8/50$. } \label{msd}
\end{center}
\end{figure}

 {We consider a network with $20$ nodes for estimating
the unknown spectrum $\boldsymbol\omega_{0}$ and set the threshold
to $\tau=1$, which according to our studies obtained the best
performance for the scenarios under evaluation}. Each iteration
corresponds to a time instant. The results are averaged over 100
experiments. The nodes scan $100$ frequencies over the frequency
axis, which is normalized between $0$ and $1$, and use $B = 50$
non$-$overlapping rectangular basis functions to model the expansion
of the spectrum \cite{dcg}. The basis functions have amplitudes
equal to one. We assume that the unknown spectrum
$\boldsymbol\omega_{0}$ is examined over 8 basis functions, leading
to a sparsity ratio equal to $S=8/50$. The power transmitted over
each basis function is set to $0.7$ mW and noise variance is set to
$0.001$. For distributed spectrum estimation, we have compared the
proposed DADMC-LMS algorithm, the oracle ATC-LMS, the RZA-ATC-LMS
\cite{lorenzo1}, the $l_0$-ATC-LMS \cite{lorenzo1} and the standard
ATC-LMS algorithms with the parameters optimized. We first measure
the performance of the algorithms in terms of MSD as shown in Fig.
\ref{msd}. The results show that DAMDC-LMS outperforms standard and
sparsity-aware algorithms and exhibits performance close to that of
the oracle algorithm, provided that the step sizes are appropriately
adjusted.

\begin{figure}[!htb]
\begin{center}
\def\epsfsize#1#2{0.85\columnwidth}
\epsfbox{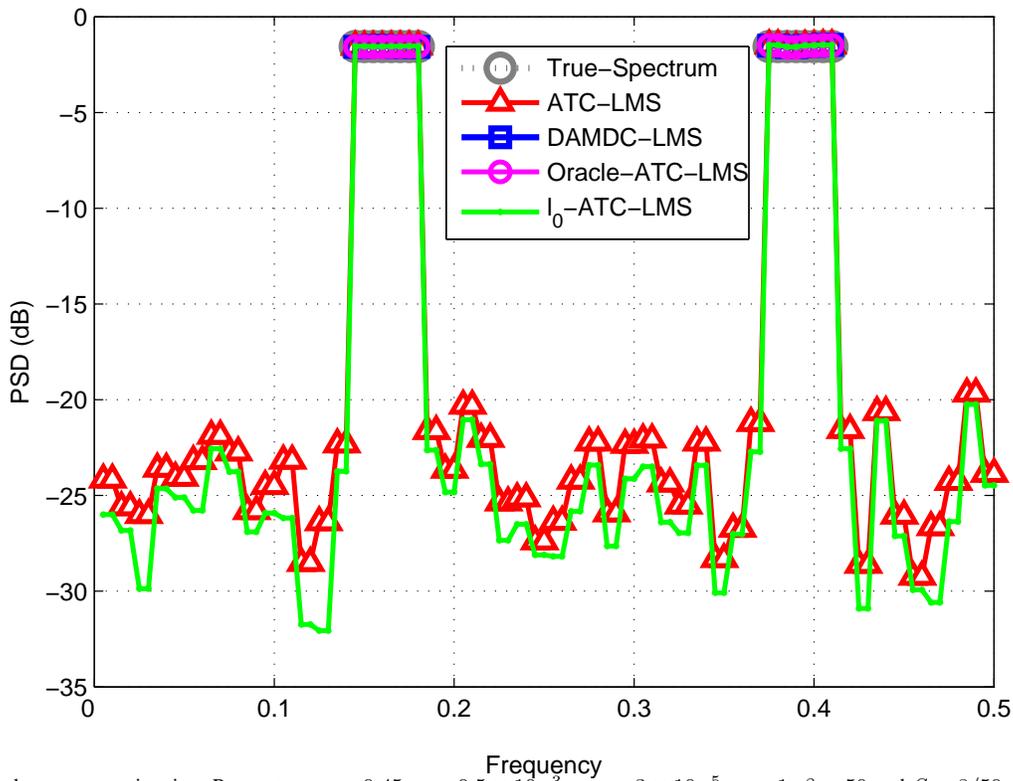} \vspace{-1.2em} \caption{Distributed spectrum
estimation. Parameters: $\mu = 0.45$, $\eta = 0.5 × 10^{-3}$
,$\rho_{l_0} = 3 × 10^{-5}$, $\tau=1$, $\beta = 50$ and $S = 8/50$.
} \label{psd}
\end{center}
\end{figure}

 {In a second example, we assess the
 {PSD} estimation performance of the algorithms. Fig.
\ref{psd} shows that the DAMDC-LMS algorithm is able to accurately
estimate the spectrum consistently with the oracle algorithm.}

\begin{figure}[!htb]
\begin{center}
\def\epsfsize#1#2{0.8\columnwidth}
\epsfbox{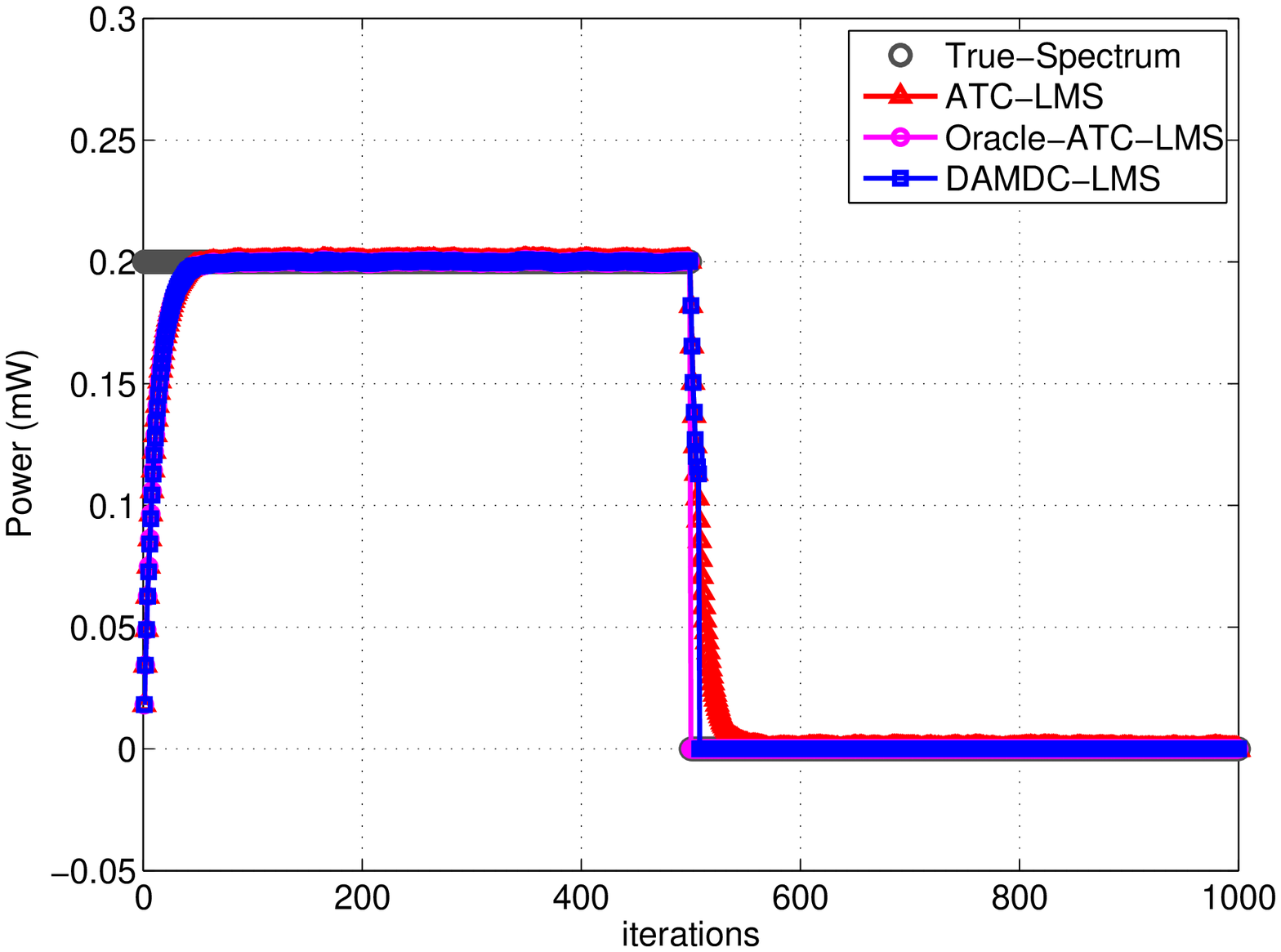} \vspace{-1.3em} \caption{\label{figure}Power
spectrum tracking. Parameters: $\mu = 0.45$, $\eta = 0.5 × 10^{-3}$
,$\rho_{l_0} = 3 × 10^{-5}$, $\beta = 50$, $\tau=1$  and $S =
8/50$.} \label{tracking}
\end{center}
\end{figure}

In order to verify the adaptation performance of DAMDC-LMS, in Fig.
\ref{tracking} we evaluate the behavior of the PSD estimates over an
initially busy channel (the $16$-th channel in this case) that
ceases to be busy after $500$ iterations, by comparing the results
achieved by the DAMDC-LMS and the oracle algorithms. We consider the
same settings of the previous example. The transmit power is set to
$0.20$ mW. We notice that DADMC-LMS is able to more effectively
track the spectrum as compared to the oracle algorithm due to its
rapid learning.

\section{Conclusion}

In this work, we have proposed a distributed sparsity-aware
algorithm for spectrum estimation over sensor networks. The proposed
DADMC-LMS algorithm outperforms previously reported algorithms.
Simulations have shown that DADMC-LMS can obtain lower MSD values
and faster convergence than prior art and close to that of the
oracle algorithm.


\newpage

\begin{thebibliography}{99}
\bibliographystyle{IEEEtran}

\bibitem{lopes}
C. G. Lopes and A. H. Sayed, ``Diffusion least-mean squares over
adaptive networks: Formulation and performance analysis,"
\textit{IEEE Transactions on Signal Processing}, vol. 56, no. 7, pp.
3122-3136, July 2008.

\bibitem{cattivelli}
F. S. Cattivelli and A. H. Sayed, ``Diffusion LMS strategies for
distributed estimation," \textit{IEEE Transactions on Signal
Processing}, vol. 58, pp. 1035-1048, March 2010.

\bibitem{mateos}
G. Mateos, J. A. Bazerque, and G. B. Giannakis, ``Distributed sparse
linear regression," \textit{IEEE Transactions on Signal Processing},
vol. 58, no. 10, pp. 5262-5276, Oct 2010.

\bibitem{candes}
E. J. Candes, M. Wakin, and S. Boyd, ``Enhancing sparsity by
reweighted l1 minimization," \textit{Journal of Fourier Analysis and
Applications}, 2008.

\bibitem{gu}
Y. Gu, J. Jin, and S. Mei, ``$L_0$-norm constraint LMS algorithm for
sparse system identification," \textit{IEEE Signal Processing
Letters}, vol. 16, pp. 774-777, 2009.

\bibitem{delamarespl1}
R. C. de Lamare and R. Sampaio-Neto, ``Adaptive reduced-rank MMSE
filtering with interpolated FIR filters and adaptive interpolators",
\textit{IEEE Sig. Proc. Letters}, vol. 12, no. 3, 2005, pp. 177 -
180.

\bibitem{delamarespl07}
R. C. de Lamare and R. Sampaio-Neto, ``Reduced-rank adaptive
filtering based on joint iterative optimization of adaptive
filters",  \textit{IEEE Signal Process. Lett.}, vol. 14, no. 12, pp.
980-983, Dec. 2007.

\bibitem{yilun}
Y. Chen, Y. Gu, and A. O. Hero, ``Sparse LMS for system
identification," \textit{Proc. IEEE International Conference on
Acoustics, Speech and Signal Processing (ICASSP)}, April 2009.

\bibitem{jidf}
R. C. de Lamare and R. Sampaio-Neto, ``Adaptive Reduced-Rank
Processing Based on Joint and Iterative Interpolation, Decimation
and Filtering", \textit{IEEE Transactions on Signal Processing},
vol. 57, no. 7, pp. 2503 - 2514, July 2009.

\bibitem{fa10}
R. Fa, R. C. de Lamare, and L. Wang, ``Reduced-Rank STAP Schemes for
Airborne Radar Based on Switched Joint Interpolation, Decimation and
Filtering Algorithm," \textit{IEEE Transactions on Signal
Processing}, vol.58, no.8, Aug. 2010, pp.4182-4194.

\bibitem{eksioglu}
E. M. Eksioglu and A. L. Tanc, ``RLS algorithm with convex
regularization," \textit{IEEE Signal Processing Letters}, vol. 18,
no. 8, pp. 470-473, August 2011.

\bibitem{angelosante}
D. Angelosante, J.A Bazerque, and G.B. Giannakis, ``Online adaptive
estimation of sparse signals: Where RLS meets the l1-norm,"
\textit{IEEE Transactions on Signal Processing}, vol. 58, no. 7, pp.
3436-3447, July 2010.

\bibitem{kalouptsidis}
N. Kalouptsidis, G. Mileounis, B. Babadi, and V. Tarokh, ``Adaptive
algorithms for sparse system identification," \textit{Signal
Processing}, vol. 91, no. 8, pp. 1910-1919, August 2011.

\bibitem{zhaocheng}
Z. Yang, R. C. de Lamare and X. Li, ``L1-Regularized STAP Algorithms
With a Generalized Sidelobe Canceler Architecture for Airborne
Radar," IEEE Transactions on Signal Processing, vol.60, no.2,
pp.674-686, Feb. 2012.

\bibitem{zhaocheng2}
Z. Yang, R. C. de Lamare and X. Li, ``Sparsity-aware space-time
adaptive processing algorithms with L1-norm regularisation for
airborne radar," IET signal processing, vol. 6, no. 5, pp. 413-423,
2012.

\bibitem{saalt}
R. C. de Lamare and R. Sampaio-Neto, ``Sparsity-aware adaptive
algorithms based on alternating optimization with shrinkage,"
\textit{IEEE Signal Processing Letters}, vol. 21, no. 2, February
2014.

\bibitem{chouvardas}
S. Chouvardas, K. Slavakis, Y. Kopsinis, and S. Theodoridis, ``A
sparsity promoting adaptive algorithm for distributed learning,"
\textit{IEEE Transactions on Signal Processing}, vol. 60, no. 10,
pp. 5412-5425, October 2012.

\bibitem{lorenzo1}
P. Di Lorenzo, S. Barbarossa and A. H. Sayed ``Distributed spectrum
estimation for small cell networks based on sparse diffusion
adaptation," \textit{IEEE Signal Processing Letters}, vol. 20, no.
12, December 2013.

\bibitem{lorenzo2} P. Di Lorenzo and S. Barbarossa,
``Distributed least-mean squares strategies for sparsity-aware
estimation over Gaussian Markov random fields," \textit{Proc. IEEE
International Conference on Acoustic, Speech and Signal Processing
(ICASSP)}, May 2014.

\bibitem{lorenzo3}
P. Di Lorenzo and A. H. Sayed, ``Sparse distributed learning based
on diffusion adaptation," \textit{IEEE Transactions on Signal
Processing}, vol. 61, no. 6, March 2013.

\bibitem{lorenzo4} P. Di Lorenzo, S. Barbarossa, and Ali H. Sayed,
``Bio-inspired decentralized radio access based on swarming
mechanisms over adaptive networks," \textit{IEEE Transactions on
Signal Processing}, Vol. 61, no. 12, pp. 3183-3197, June 2013.

\bibitem{arablouei}
R. Arablouei, S. Werner, Y.-F. Huang and K. Dogançay, ``Distributed
least mean-square estimation with partial diffusion," \textit{IEEE
Transactions on Signal Processing}, vol. 62, No. 2, pp. 472-484,
January 2014.

\bibitem{liu}
Z. Liu, Y. Liu and C. Li, ``Distributed sparse recursive
least-squares over networks," \textit{IEEE Transactions on Signal
Processing}, vol. 62, no. 6, pp. 1386-1395, March 2014.

\bibitem{dcg} S. Xu and R. C. de Lamare, ``Distributed
conjugate gradient strategies for distributed estimation over sensor
networks," \textit{Proc. Sensor Signal Processing for Defense
(SSPD)}, September 2012

\bibitem{dce}
S. Xu, R. C. de Lamare and H. V. Poor, ``Distributed compressed
estimation based on compressive sensing," \textit{IEEE Signal
Processing Letters}, vol. 22, no. 9, September 2014.

\bibitem{dta_ls}
S. Xu, R. C. de Lamare and H. V. Poor, ``Adaptive link selection
algorithms for distributed estimation," \textit{EURASIP Journal on
Advances in Signal Processing}, 2015.

\bibitem{dcg_iet}
S. Xu, R. C. de Lamare and H. V. Poor, ``Distributed estimation over
sensor networks based on distributed conjugate gradient strategies,"
\textit{IET Signal Processing}, 2016.

\bibitem{vardhan}
K. Vardhan, S. Mohammed, A. Chockalingam, and B. Rajan, ``A
low-complexity detector for large MIMO systems and multicarrier CDMA
systems," \textit{IEEE Journal on Selected Areas in Communications},
vol. 26, no. 3, p. 473-485, 2008.

\bibitem{las_li}
P. Li and R. C. Murch, ``Multiple output selection-LAS algorithm in
large MIMO systems," \textit{IEEE Communications Letters}, vol.14,
no.5, pp.399-401, May 2010.

\bibitem{ref12} O. Axelson,
\textit{Iterative Solution Methods}, Cambridge Univ. Press, 1994.

\bibitem{ref13}
G. H. Golub and C. F. Van Loan, \textit{Matrix Computations}, 2nd
Ed. Baltimore, MD: Johns Hopkins Univ. Press, 1989.

\bibitem{ref14}
S. Theodoridis, \textit{Machine Learning: a Bayesian and
Optimization Perspective}, Academic Press, March 2015.

\bibitem{ref15}
N. A. Lynch, \textit{Distributed Algorithms}, Morgan Kaufmann, 1997.

\bibitem{ref15}
O. Jahromi and P. Aarabi, ``Distributed spectrum estimation in
sensor networks," \textit{Proc. IEEE International Conference on
Acoustics, Speech, and Signal Processing}, vol. 3, pages. 849-52,
May 2004.

\end{thebibliography}
\end{document}